Scientific machine learning in Hydrology: a unified perspective

**Authors:** Adoubi Vincent De Paul Adombi[1,*]

1. Research Group R2Eau, Centre d'études sur les ressources minérales, Université du Québec à Chicoutimi, 555 boulevard de l'Université, Chicoutimi, Québec G7H 2B1, Canada. **Email\*: avadombi@uqac.ca** | ORCID id: 0000-0001-8063-6444

**Abstract**

Scientific machine learning (SciML) provides a structured approach to integrating physical knowledge into data-driven modeling, offering significant potential for advancing hydrological research. In recent years, multiple methodological families have emerged, including physics-informed machine learning, physics-guided machine learning, hybrid physics–machine learning, and data-driven physics discovery. Within each of these families, a proliferation of heterogeneous approaches has grown up independently, often lacking conceptual coordination. This fragmentation complicates the assessment of methodological novelty and makes it difficult to identify where meaningful advances can still be made in the absence of a unified conceptual framework. This review, the first focused overview of SciML in hydrology, addresses these limitations by proposing a unified methodological framework for each SciML family, bringing together representative contributions into a coherent structure that fosters conceptual clarity and supports cumulative progress in hydrological modeling. Finally, the limitations and future opportunities of each unified family were highlighted to guide systematic research in hydrology, where these methods remain underutilized.

## 1. Introduction

Scientific Machine Learning (SciML) is a general paradigm that integrates physical knowledge with machine learning modeling, thus providing a transformative approach to advancing hydrological research and addressing the complex challenges of water systems (Cuomo et al., 2022; Thiyagalingam et al., 2022). Its recent development has been motivated by longstanding challenges in the field such as the limited and noisy observational data that hinder the development of robust traditional machine learning models (Chen et al., 2021b; Satyadharma et al., 2024; Yang et al., 2021), the limited accuracy and high computational cost of traditional numerical simulators (Chen et al., 2020; Di Salvo, 2022), the need for greater flexibility in representing complex and heterogeneous system dynamics (Meng et al., 2025; Raissi et al., 2019), the persistent gap between physical interpretability and predictive performance (Biazar et al., 2025; Sang, 2013), and the increasing demand for data-driven discovery of new processes or refinement of existing physical knowledge (Shu and Ye, 2023; Wang et al., 2023).

Several methodological families have emerged under the SciML umbrella, including physics-informed machine learning (Huang et al., 2023; Ji et al., 2024), hybrid physics–ML models (Okkan et al., 2021; Roy et al., 2023), physics-guided machine learning (Chen et al., 2022; Chen et al., 2023b; Palmitessa et al., 2022), and data-driven approaches for physics discovery (Brunton and Kutz, 2019; Floryan and Graham, 2022; North et al., 2023). Each reflects a different strategy for incorporating domain knowledge.

Over the course of the development of the literature, these methodological families have evolved in parallel, producing a proliferation of methods adapted to specific tasks or application contexts.

However, many of these contributions are presented in isolation (e.g., Xu et al., 2023), with limited reference to common principles or shared design structures. This fragmentation not only complicates the assessment of methodological novelty and makes it difficult to identify areas where significant progress can still be made in the absence of a unified conceptual framework, it also raises the barrier of entry for newcomers to the field and slows down interdisciplinary exchanges. Researchers with a background in machine learning may struggle to identify the relevant modeling principles specific to hydrology, while hydrologists may find it difficult to situate new contributions within the broader methodological landscape. The lack of abstraction thus limits both accessibility and cumulative progress.

This review fills that gap as the first dedicated synthesis of SciML in hydrology. It outlines a methodological guide structured around the four families of approaches mentioned above. Rather than striving for exhaustiveness, we highlight representative contributions that reveal core modeling strategies and recurring design patterns. The aim is to provide a unified conceptual perspective that clarifies the internal logic of each family, fosters clearer communication, encourages methodological reuse, and lowers the conceptual entry barrier to this rapidly evolving field. The remainder of the paper is organized as follows: for each family of methods, we first present the unified framework, then illustrate how it is instantiated through representative case studies, and finally discuss associated limitations and future opportunities.

## 2. Unified physics-informed machine learning (UPIML)

Physics-informed machine learning (PIML) designates a broad family of approaches that incorporate known physical laws (e.g., governing equation, initial and boundary conditions) as constraints within the training objective of machine learning models (Karniadakis et al., 2021; Raissi et al., 2019). This integration ensures that the learned model respects fundamental physical principles, improving generalization, interpretability, and robustness, especially when observational data are limited or noisy.

To systematize and unify the broad family of physics-informed models, we propose the unified physics-informed machine learning (UPIML) framework. UPIML formalizes the key components common to many PIML approaches in hydrology, providing a modular architecture alongside a composite loss function that explicitly integrates multiple forms of physical and data-driven constraints.

## 2.1. Modular architecture of UPIML and composite loss function

UPIML is structured into two main types of modules, each handling different parts of the physical modeling process: parameterization modules and state modules as illustrated in **Fig. 1**.

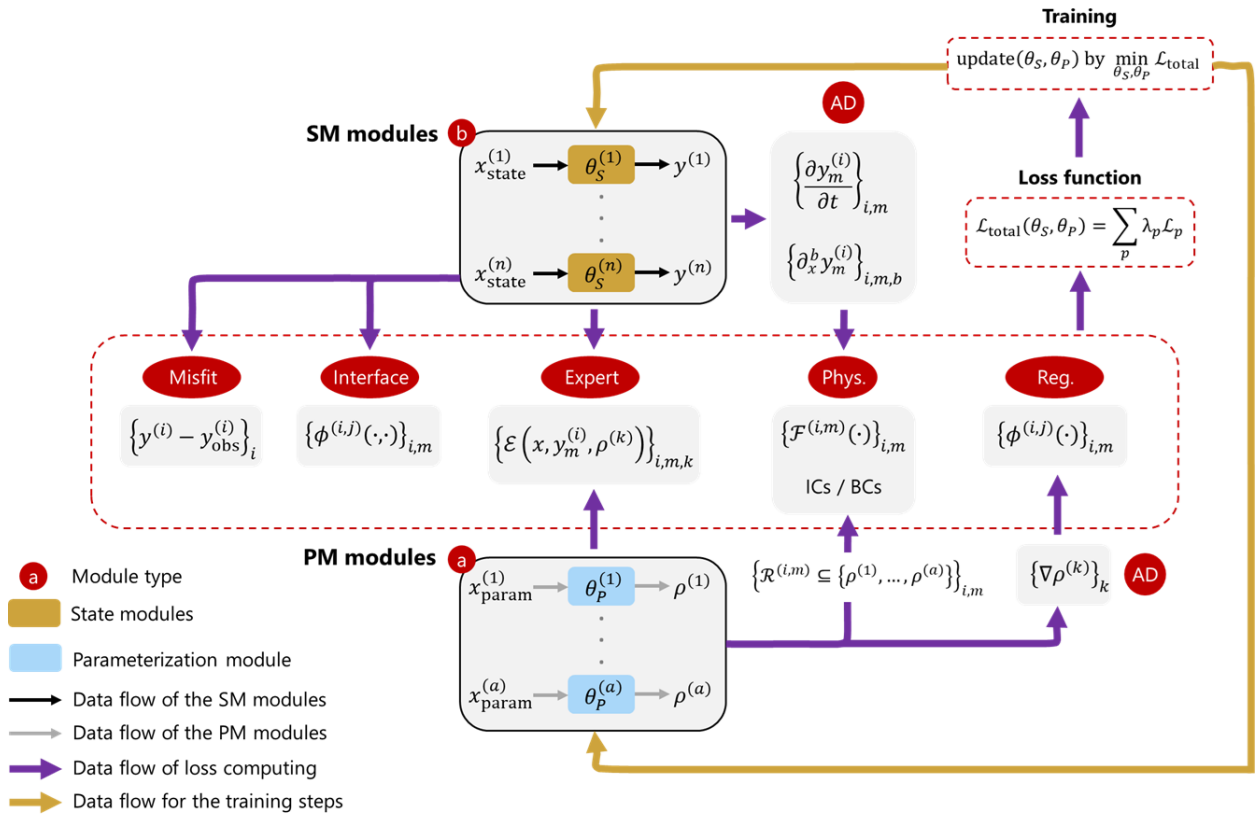

**Fig. 1** Architecture of the Unified Physics-Informed Modular Network (UPIML). The framework consists of two types of modules: State Modules (SMs) (gold) that predict physical state variables $y^{(i)}$, and Parameterization Modules (PMs) (blue) that estimate latent or physical parameter fields

$\rho^{(k)}$. Arrows indicate data flow: within modules (black and gray), toward loss computation (purple), and for parameter updates (orange). Loss components include misfit to observations, interface continuity, expert-based constraints, physics-informed residuals (using automatic differentiation, AD), and regularization on parameter fields. These terms are combined in a global loss $\mathcal{L}_{total}$, minimized jointly over module parameters $\theta_S = \left(\theta_S^{(1)}, \ldots, \theta_S^{(n)}\right)$ and $\theta_P = \left(\theta_P^{(1)}, \ldots, \theta_P^{(a)}\right)$ during training.

### 2.1.1. Parameterization modules (PMs)

Each PM estimates spatially distributed latent or physical parameters such as hydraulic conductivity or porosity based on auxiliary inputs like soil characteristics, land cover, or topographic descriptors (Luo et al., 2024). An UPIML architecture may incorporate one or more PMs, depending on the complexity and heterogeneity of the modeled system. Let $\text{PM}^{(k)}$ denote the $k$-th parameterization module, typically implemented as a machine learning model (e.g., a neural network). Let $x_{\text{param}}^{(k)} \in \mathbb{R}^{d_k}$ represent its input features of dimensionality $d_k$, and $\rho^{(k)} \in \mathbb{R}^{p_k}$ the resulting predicted parameter field of dimensionality $p_k$. $\rho^{(k)}$ is given by **Eq. 1**.

$$\rho^{(k)}(x) = \text{PM}^{(k)}\left(x_{\text{param}}^{(k)}, \theta_P^{(k)}\right) \qquad 1$$

where $\theta_P^{(k)}$ are the learnable parameters of the $k$-th module.

### 2.1.2. State modules (SMs)

Each state module (SM) predicts physical state variables such as pressure, velocity, water level, or water discharge given input features like spatial location, time, and environmental forcings (Bandai and Ghezzehei, 2022). These outputs can be scalar or vector-valued, depending on the modeled process. Let $SM^{(i)}$ denotes the $i$-th state module, typically implemented as a neural network or another differentiable function approximator. Let $x_{\text{state}}^{(i)} \in \mathbb{R}^{q_i}$ denote the input vector of dimensionality $q_i$ that may include time, spatial coordinates, and other external features, and let $y^{(i)} \in \mathbb{R}^{s_i}$ be the predicted physical state vector with $s_i$ output components. $y^{(i)}$ is given by **Eq. 2**.

$$y^{(i)}(x) = SM^{(i)}\left(x_{\text{state}}^{(i)}, \theta_S^{(i)}\right) \qquad 2$$

where $\theta_S^{(i)}$ are the learnable parameters of the $i$-th state module.

### 2.1.3. Composite loss function

The total loss $\mathcal{L}_{\text{total}}$ in an UPIML model combines multiple components to balance observational fidelity, physical consistency, inter-module coherence, expert knowledge, and regularization as given by **Eq. 3**.

$$\mathcal{L}_{\text{total}} = \lambda_{\text{misfit}}\mathcal{L}_{\text{misfit}} + \lambda_{\text{phys}}\mathcal{L}_{\text{phys}} + \lambda_{\text{interface}}\mathcal{L}_{\text{interface}} + \lambda_{\text{expert}}\mathcal{L}_{\text{expert}} \\ + \lambda_{\text{reg}}\mathcal{L}_{\text{reg}} \qquad 3$$

where each $\lambda \geq 0$ is a tunable weight that controls the relative contribution of each term.

- **Data misfit loss $\mathcal{L}_{\text{misfit}}$**

This term enforces agreement between model predictions and observed data through supervised learning (**Eq. 4**).

$$\mathcal{L}_{\text{misfit}} = \sum_{i=1}^{n} \sum_{x \in \mathcal{D}_{\text{obs}}^{(i)}} \left\| y^{(i)}(x) - y_{\text{obs}}^{(i)}(x) \right\|^2 \qquad 4$$

where $y^{(i)} \in \mathbb{R}^{S_i}$ is the model prediction from state module $i = 1, \dots, n$; $y_{\text{obs}}^{(i)} \in \mathbb{R}^{S_i}$ is the observed counterpart at $x$, $\mathcal{D}_{\text{obs}}^{(i)}$ denotes the set of observed points used for the $i$-th module; $\|\cdot\|$ denotes, for example, the Euclidean norm.

- **Physics-based loss $\mathcal{L}_{\text{phys}}$**

To ensure that the predictions produced by the UPIML are physically consistent, we incorporate explicit constraints derived from governing physical laws (e.g., conservation equations, constitutive relationships) into the loss function. This is done through the physics-based loss term $\mathcal{L}_{\text{phys}}$, which penalizes violations of these laws over a prescribed domain. Each of the output variables of the multivariate prediction $y^{(i)}$ is typically governed by a distinct partial differential equation (PDE). Therefore, we define the PDE residual $\mathcal{F}^{(i,m)}$ associated to the $m$-th output component $y_m^{(i)}$ as given by **Eq. 5**.

$$\mathcal{F}^{(i,m)}\left( y_m^{(i)}(x), \left\{ \partial_x^b y_m^{(i)}(x) \right\}_{i,m,b}, \mathcal{R}^{(i,m)}(x) \right) = 0 \qquad 5$$

where $\partial_x^b y_m^{(i)}(x)$ denotes the $b$-th spatial (or temporal) derivative of $y_m^{(i)}$; and can be computed automatically via automatic differentiation (AD), enabling seamless integration with modern deep learning frameworks; $\mathcal{R}^{(i,m)}(x) \subseteq \{\rho^{(1)}(x), \dots, \rho^{(k)}(x), \dots, \rho^{(a)}(x)\}$ denotes the set of parameter fields used in the formulation of the residual $\mathcal{F}^{(i,m)}$. This set may involve only a subset of the predicted parameters, possibly coming from one or several parameterization modules.

Since PINNs, and by extension UPIMLs, may violate temporal causality during training, potentially leading to spurious or non-physical solutions, Wang et al. (2024) proposed a mitigation strategy based on a weighted physical residual loss. At a given time step $t_k$, we define $\mathcal{L}_{\text{phys}}^{(i,m)}(t_k)$ as the contribution to the physics-based loss for the pair (state module $i$, output $m$), as formalized in **Eq. 6**.

$$\mathcal{L}_{\text{phys}}^{(i,m)}(t_k) = \sum_{x \in \mathcal{D}_{\text{phys},k}^{(i,m)}} \left\| \mathcal{F}^{(i,m)}\left( y_m^{(i)}(x), \left\{ \partial_x^b y_m^{(i)}(x) \right\}_{i,\,m,\,b}, \mathcal{R}^{(i,m)}(x) \right) \right\|^2 \quad \quad 6$$

The physics-based loss is then given by **Eq. 7**.

$$\mathcal{L}_{\text{phys}} = \sum_{i=1}^{n} \sum_{m=1}^{s_i} \left[ \frac{1}{N_t^{(i,m)}} \sum_{k=1}^{N_t^{(i,m)}} w_k^{(i,m)} \mathcal{L}_{\text{phys}}^{(i,m)}(t_k) \right] \quad \quad 7$$

where $\mathcal{D}_{\text{phys},k}^{(i,m)}$ is the set of spatial (and possibly spatio-temporal) locations where the residual $\mathcal{F}^{(i,m)}$ is enforced for the $m$-th output variable of module $i$. Individual causal weights $w_k^{(i,m)}$ are defined by **Eq. 8**.

$$w_k^{(i,m)} = \exp\left( -\epsilon \sum_{j=1}^{k-1} \mathcal{L}_{\text{phys}}^{(i,m)}(t_j) \right) \quad \quad 8$$

Where $N_t^{(i,m)}$ represents the total number of time steps at which the PDE residual is evaluated; $\epsilon$ is a causality parameter that controls the steepness of the weights. Such a loss function allows $\mathcal{L}_{\text{phys}}^{(i,m)}(t_k)$ to be minimized only if all residues $\mathcal{L}_{\text{phys}}^{(i,m)}(t_j)$ before $t_k$ are correctly minimized, and vice versa. Note that boundary and initial conditions can be included as hard constraints directly encoded in the neural network design or as soft constraints via penalties in $\mathcal{L}_{\text{phys}}$. An example of soft constraint of a Dirichlet boundary condition can be given by **Eq. 9**.

$$\mathcal{L}_{\text{BC}} = \sum_{i=1}^{n} \sum_{x \in \partial \Omega} \left\| y_m^{(i)}(x) - g(x) \right\|^2 \qquad 9$$

where $g(x)$ is a known target value that the model output is expected to match on the boundary $\partial \Omega$.

- **Interface loss $\mathcal{L}_{\text{interface}}$**

The interface loss term encourages coherence between interconnected state modules (e.g., mass flux continuity, or pressure–velocity coupling) and it is given by **Eq. 10**.

$$\mathcal{L}_{\text{interface}} = \sum_{(i,j) \in \mathcal{C}} \sum_{x \in \Gamma_{ij}} \left\| \phi^{(i,j)} \left( y^{(i)}(x), y^{(j)}(x) \right) \right\|^2 \qquad 10$$

where $\mathcal{C}$ is the set of coupled module pairs; $\Gamma_{ij}$ denotes the spatial interface between modules $i$ and $j$; $\phi^{(i,j)}$ is a user-defined function encoding the interface constraint.

- **Expert knowledge loss $\mathcal{L}_{\text{expert}}$**

The expert knowledge loss term encodes empirical relationships, inequality constraints, or heuristic knowledge, often unavailable in formal physics (**Eq. 11**).

$$\mathcal{L}_{\text{expert}} = \sum_{i=1}^{n} \sum_{m=1}^{s_i} \sum_{x \in \mathcal{D}_{\text{expert}}} \left\| \mathcal{E} \left( x, y_m^{(i)}(x), \rho^{(k)}(x) \right) \right\|^2 \qquad 11$$

where $\mathcal{E}$ is a function of one or a combination of its variables and may include rules such as admissible parameter ranges or monotonic relationships. For example, for hydraulic conductivity $\rho^{(1)}(x) \in [10^{-6}, 10^{-2}]$, a hinge loss can be defined by **Eq. 12**.

$$\mathcal{E}\left(x, \rho^{(1)}(x)\right) = \max(0, \rho^{(1)}(x) - 10^{-2}) + \max(0, 10^{-6} - \rho^{(1)}(x)) \qquad 12$$

- **Regularization loss $\mathcal{L}_{reg}$**

This term promotes generalization and smoothness in learned fields and weights. It can be defined by **Eq. 13**.

$$\mathcal{L}_{reg} = \sum_{k=1}^{a}\left(\max(0, \|\nabla \rho^{(k)}(x)\| - \tau)\right)^2 + \sum_{k=1}^{a}\left\|\theta_P^{(k)}\right\|^2 + \sum_{i=1}^{n}\left\|\theta_S^{(i)}\right\|^2 \qquad 13$$

In this loss, the first term penalizes the spatial roughness of the parameter fields when their variation is greater than a threshold $\tau$, while the other terms apply a weight decay to reduce overfitting in the parameterization and state modules.

### 2.1.4. Sampling strategies for residual evaluation

To evaluate loss components that depend on spatial and/or temporal variables such as the physics-based residual loss or boundary condition loss, it is necessary to sample specific space-time points within the study domain. These evaluation points are commonly referred to as collocation points, i.e., the locations where physical constraints (e.g., PDE residuals) are enforced. While early PINN studies often relied on simple sampling strategies like equispaced grids or uniform random distributions (Wu et al., 2023), more sophisticated techniques have emerged. For instance, low-discrepancy sequences such as Sobol or Halton (Halton, 1960; Sobol, 1967) have been used to improve point coverage. Adaptive sampling strategies have also shown promise: by selecting collocation points that contribute the most informative gradients (Nabian et al., 2021), or by refining point distribution based on residual errors (Wight and Zhao, 2020; Wu et al., 2023), these approaches accelerate convergence and improve solution accuracy. Time-adaptive methods, which either focus sampling in critical temporal regions or segment time into smaller subdomains, have further enhanced performance in problems with sharp dynamics (Wight and Zhao, 2020). Overall, the choice of collocation point sampling strategy can significantly affect training efficiency and model fidelity, especially in complex or high-dimensional problems.

### 2.2. Special cases of UPIML

In this section, we demonstrate that by appropriately constraining the UPIML architecture (e.g. omitting certain modules), simplifying the loss formulation or fixing parts of the parameter or state representations, existing approaches in the literature can be reinterpreted as specific instantiations within the broader UPIML framework. This highlights the unifying nature of UPIML and its ability to generalize a wide range of physics-informed modeling strategies.

#### 2.2.1. Single-state and no-parameterization architectures

In this first case study, we demonstrate how classical PINN-based forward modeling approaches, commonly encountered in the hydrological literature, can be recovered as simplified instances of UPIML. These configurations typically involve a single state module with no parameterization module and use a relatively simple additive loss function that may or may not incorporate domain-specific expert knowledge.

For example, Ali et al. (2024) use a single neural network that maps space-time inputs $(t, x)$ to the groundwater level $\phi(t, x)$. This corresponds, in UPIML terms, to a configuration where only one state module is instantiated, no parameter module is present, physical parameters are assumed known or constant. The loss function consists of a sum of mean squared errors on the PDE residual, initial and boundary conditions, each unweighted $(\lambda_i = 1)$. This setup is equivalent to a basic UPIML where the architectural complexity is minimized, and no adaptive inference of physical parameters is involved.

In a related application, Nazari et al. (2024) consider water flow in a river channel and adopt a similar one-network architecture. However, they augment the standard PINN loss with two expert-informed regularization terms: a penalty that enforces the physical parameters (Manning coefficient and bed slope) to remain close to prior means, akin to adding a parameter trust region; a penalty enforcing agreement between the time derivative of water level computed by automatic differentiation and the same derivative by numerical method. This formulation still aligns with the UPIML framework, using one state module, a fixed or weakly inferred parameter vector, again without a dedicated parameterization module, and an augmented loss that combines physics, data and expertise constraints with equal weights.

Finally, Dazzi (2024) extend this line of work by solving the augmented shallow water equations with topography. Their approach also relies on a single neural network but further incorporates domain-specific hard or soft constraints, such as enforcing non-negativity of the water depth $h \geq 0$, zero velocity in dry regions, and differentiated weighting for various components of the loss function.

### 2.2.2. Multi-state modules via domain decomposition or multi-physics modeling

This second case study illustrates how UPIML recovers PINN architecture based on multiple state modules, a structure encountered in both domain decomposition and multiphysics contexts.

For example, Bandai and Ghezzehei (2022) addressed the simulation of unsaturated flow in layered soils by solving the 1D Richardson-Richard equation. In this setting, the physical domain is divided into two subdomains, each corresponding to a distinct soil layer with potentially different hydraulic properties. UPIML accommodates this structure through multiple state modules (one per layer), each responsible for learning the solution in its corresponding subdomain. No parameterization module is included. Continuity conditions at the interface between the two subdomains are enforced via additional loss terms that penalize discrepancies in the physical state and associated fluxes between the two state modules. This setup fits directly within the UPIML framework as a special case involving a multi-state configuration with interface consistency constraints integrated into the loss.

A different use of multiple state modules is found in Haruzi and Moreno (2023), who proposed a multiphysics-informed PINN to solve a 2D transient flow and solute transport problem in the unsaturated zone. The architecture consists of two distinct neural networks, each responsible for approximating one of the coupled physical fields (hydraulic head and solute concentration). Training is performed using geoelectrical measurements, and the model is designed to operate under incomplete boundary conditions and missing initial condition. As with the previous study, this configuration maps directly to UPIML using multiple state modules, no explicit parameterization, and a customized loss structure.

### 2.2.3. Multi-state modules with parameterization module

This case illustrates how UPIML recovers architectures combining multiple state modules, a parameterization module, and embedded expert constraints. Luo et al. (2024) modeled river network hydrodynamics using an enhanced PINN architecture with the following features: (i) two state modules: one predicting water surface elevation from time and space coordinates, and another predicting flow velocity from the same inputs; (ii) a pretrained parameterization module that maps the predicted elevation to hydraulic properties such as flow area, river width, and conveyance; (iii) expert knowledge loss terms, enforcing physical constraints like mass conservation at junctions and bounds on water surface elevation; (iv) adaptive loss weighting to balance different components of the training objective.

## 2.3. Limitations and opportunities of UPIML

The application of UPIML in hydrology remains limited by several technical obstacles. A major limitation is the high computational cost of training these models, which often significantly exceeds that of traditional numerical models (Ali et al., 2024; Bandai and Ghezzehei, 2022; Luo et al., 2024). To mitigate this issue, recent studies have explored parallelization strategies. For instance, Meng et al. (2020) introduced the parareal PIML, which decomposes long-time simulations into multiple short-time subproblems that can be solved independently and supervised by a coarse-grained solver. Similarly, Jagtap et al. (2020) developed a conservative PIML as a promising approach to reduce computational overhead due to its parallelization capacity even if it has not been implemented in their study. Consequently, practical and scalable parallelization in UPIML remains an open research challenge. As with traditional machine learning models, UPIML approaches are susceptible to underfitting/overfitting (Ali et al., 2024; Zou et al., 2024) and are highly sensitive to neural network initialization (Bandai and Ghezzehei, 2022). Since UPIMLs are based on standard ML models trained to minimize a composite loss function incorporating physical constraints, once training is complete, the loss terms are no longer active. This decouples the model from the physical laws governing it. As a result, under changing conditions such as modifications in boundary conditions or the introduction of a pumping well in an aquifer system, the trained model often fails to generalize and then behaves like a static PDE solver. An important opportunity lies in the development of UPIML architectures that retain an explicit and dynamic representation of the underlying physics. This will enable adaptation to new scenarios without re-training the model, thus improving model transferability. An example of this type of architecture is the PDE-preserved neural network (PPNN) (Liu et al., 2024), which integrates discretized forms of the governing equations directly into the neural network structure. Finally, it should be noted that UPIML may only guarantee a local optimum, with low solution accuracy, which may limit its generalizability (Zhang et al., 2024).

## 3. Unified physics-guided machine learning (UPGML)

Unlike UPIML, which enforces physical consistency through constraints in the loss function, PGML exploits physical signals usually derived from deterministic or stochastic models as features, intermediate representations or auxiliary variables to inductively guide the learning process. To systematize this broad family of models, the unified physics-guided machine learning (UPGML) framework is presented. UPGML formalizes the key components shared by the PGML methods in hydrology, providing a modular architecture where physical and exogenous data interact at several stages of the predictive pipeline.

### 3.1. Modular architecture of UPGML

UPGML is structured into five modules, each handling different parts of the physical modeling process: physical feature generation, input encoding, latent representation learning, latent fusion and output mapping as illustrated in **Fig. 2**.

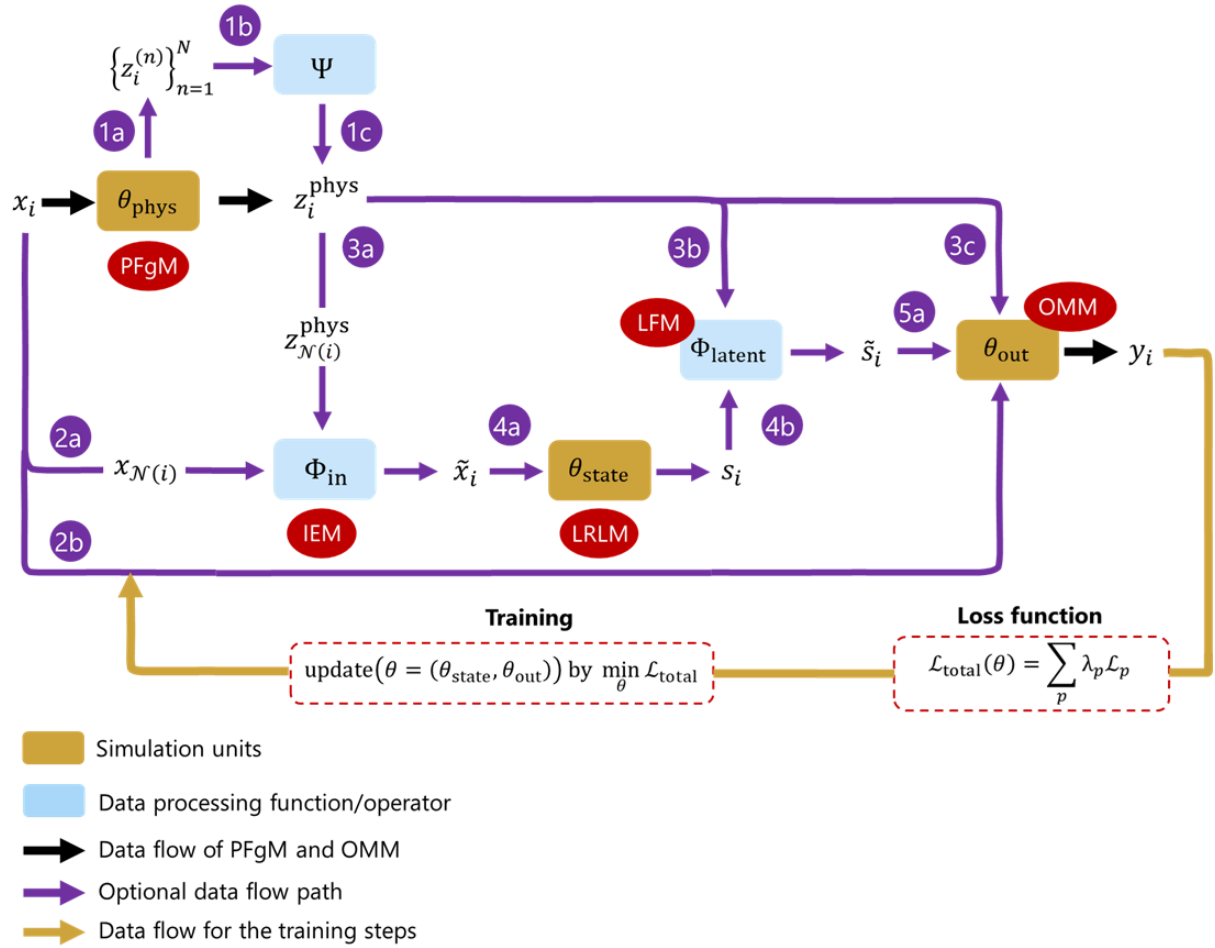

**Fig. 2** Architecture of the unified physics-guided machine learning framework (UPGML). The architecture integrates modular components that combine physical simulation with data-driven learning for hydrological prediction. It is composed of two main types of blocks: simulation units (gold) that perform deterministic or stochastic physical simulations and predictive mapping and data processing functions/operators (blue) that transform and integrate inputs at various levels of abstraction. Each of these blocks is labeled with a red module identifier indicating its role in the overall pipeline: physical feature generation module (PFgM), input encoding module (IEM), latent representation learning module (LRLM), latent fusion module (LFM), and output mapping module (OMM). The full predictive pipeline processes exogenous forcings $x_i$ through the PFgM to generate physical signals $z_i^{phys}$ which are aggregated via $\Psi$ in the stochastic case and combined

with raw inputs via the operator $\Phi_{in}$. A compact latent representation $s_i$ is then learned by the LRLM and may be fused mid-network with physical features via $\Phi_{latent}$. Final predictions $y_i$ are computed using the OMM from the enriched state $\tilde{s}_i$. Arrows indicate the flow of information: black arrows denote the primary forward path from physical input to prediction; purple arrows represent optional or intermediate data flows, including physical injection into latent space; and orange arrows denote the flow of gradients and data used during training. The model is trained end-to-end by minimizing a composite loss $\mathcal{L}_{total}(\theta) = \sum_p \lambda_p \mathcal{L}_p$, where each component $\mathcal{L}_p$ include data misfit, regularization, or physically motivated constraints. Parameters $\theta = (\theta_{state}, \theta_{out})$ are jointly optimized during learning.

Here, the objective is to estimate a target hydrological variable $y_i \in \mathbb{R}^m$ (e.g., streamflow) from a set of exogenous forcings $x_i \in \mathbb{R}^n$ (e.g., precipitation, evapotranspiration) over a set of indices $I$ (e.g., time steps $t = 1, \ldots, T$). In many cases, the value of $y_i$ should depend not only on the current forcing $x_i$, but also on a finite temporal or spatial neighborhood $\mathcal{N}(i) \subseteq I$.

### 3.1.1. Physical feature generation module (PFgM)

The UPGML architecture begins with the physical feature generation module (PFgM). This module incorporates prior scientific knowledge into the learning process in the form of outputs from a physics-based simulator $F_{\text{phys}}$ using the exogenous forcings (**Eq. 14**).

$$z_i^{\text{phys}} = F_{\text{phys}}(x_{\mathcal{N}(i)}, \theta_{\text{phys}}) \qquad 14$$

where $\theta_{\text{phys}}$ denotes either known or calibrated physical parameters. The physics-based simulator can be, for example, a numerical model (Kim et al., 2008; Sahu et al., 2023). In cases involving physics-based stochastic models, as in Houénafa et al. (2025), the physics-based output may consist of a sample of trajectories $\{z_i^{(n)}\}_{n=1}^{N} \sim F_{\text{phys}}^{\text{stoch}}$ to which an aggregation function $\Psi$ (e.g. mean, variance) is applied to obtain a "deterministic" representation (**Eq. 15**).

$$z_i^{\text{phys}} = \Psi\left(\{z_i^{(n)}\}_{n=1}^{N}\right) \qquad 15$$

This module establishes the foundation for physical guidance within the UPGM architecture.

### 3.1.2. Input encoding module (IEM)

The second stage of the architecture, represented by the input encoding module (IEM), combines the physical outputs produced in the PFgM module with the raw exogenous forcings via an input encoding function $\Phi_{\text{in}}$ to build a unified, information-rich input representation $\tilde{x}_i$ (**Eq. 16**).

$$\tilde{x}_i = \Phi_{\text{in}}\left(x_{\mathcal{N}(i)}, z_{\mathcal{N}(i)}^{\text{phys}}\right) \in \mathbb{R}^d \qquad 16$$

where $\Phi_{\text{in}}$ ranges from raw concatenation (Lu et al., 2021; Young et al., 2017) to sophisticated signal decompositions via wavelet transforms (Houénafa et al., 2025).

### 3.1.3. Latent representation learning module (LRLM)

The encoded inputs $\tilde{x}_i$ are then fed to a latent representation learning module (LRLM), typically implemented as a neural network (e.g., MLP, LSTM) (Feng et al., 2024; Widiasari et al., 2017), to capture latent dynamics $s_i$ of the hydrological system (**Eq. 17**).

$$s_i = F_{\text{state}}(\tilde{x}_i, \theta_{\text{state}}) \in \mathbb{R}^q \qquad 17$$

where $\theta_{\text{state}}$ denotes the LRLM module parameters. This module can be used either to directly predict target variables such as streamflow, or to generate intermediate representations that abstract key features of the system's behavior.

### 3.1.4. Latent fusion module (LFM)

In some architectures, a latent fusion module (LFM) is used to incorporate physical information at deeper levels of the machine learning model via latent fusion mechanisms. This enables physical guidance in the core of the model architecture and improved interpretability. To formalize this, a latent fusion operator $\Phi_{\text{latent}}$ is defined, which integrates physical features with the learned latent state $s_i$ to obtain a fused latent state $\tilde{s}_i$ (**Eq. 18**).

$$\tilde{s}_i = \Phi_{\text{latent}}\left(s_i, z_i^{\text{phys}}\right) \qquad 18$$

If LFM is omitted, this defaults to $\tilde{S}_i = S_i$. $\Phi_{\text{latent}}$ can adopt many forms, such as a simple injection of data into the neurons of a hidden layer of a neural network, concatenation, feature modulation, projection or an attention mechanism (Niu et al., 2021).

### 3.1.5. Output mapping module (OMM)

The output mapping module (OMM) provides the final predictive output, calculated from the fused latent state using an output transformation, as shown in **Eq. 19**.

$$y_i = F_{\text{out}}(\tilde{s}_i, \theta_{\text{out}}) \in \mathbb{R}^m \qquad 19$$

where $\theta_{\text{out}}$ denotes the OMM module parameters.

## 3.2. Special cases of UPGML

### 3.2.1. Input-level hybridization

In the first case, the outputs of physics-based simulations are directly used as additional inputs to a machine learning model, which then predicts the target variable. This shallow fusion mechanism corresponds to an UPGML configuration in which the physics module (PFgM) generates observable quantities $z_i^{\text{phys}}$, which are concatenated to the raw inputs $x_i$ (paths 2a-3a and IEM, see **Fig. 2**) and passed on to a predictive ML model (LRLM) $F_{\text{state}}$. The loss function is exclusively formulated based on discrepancies between model predictions and observed data, without the inclusion of any physics-informed constraints. This configuration is used by Young et al. (2017), who integrate the outputs of the physics-based HEC-HMS model along with meteorological forcings into machine learning models, in particular a feedforward neural network and a support vector machine, to predict hourly runoff. Similarly, Lu et al. (2021) integrate outputs from physics-based hydrological simulations with meteorological forcings and use the combined features as input to an LSTM network for daily streamflow prediction in data-scarce regions.

### 3.2.2. Statistical feature extraction from stochastic physics-based model

The second special case involves the extraction of time-dependent statistical summaries from a physics-based stochastic simulator. These summaries, such as the mean and variance of simulated discharge, are transformed and used as input features to the ML model. The fusion between physics and machine learning still occurs at the input level. This approach is employed by Houénafa et al. (2025), who use a stochastic rainfall-runoff model to generate an ensemble of discharge trajectories from meteorological inputs (path 1a, **Fig. 2**). For each time step, statistics such as the mean discharge are computed over this ensemble, then preprocessed using a wavelet transform and optionally augmented with time-lagged values (paths 1b-1c-3a, **Fig. 2**). The resulting features are passed to a gated recurrent unit (GRU) model to predict discharge (path 4a, **Fig. 2**).

### 3.2.3. Latent fusion of physics and data representations

The third special case leverages latent fusion, wherein physics-based information is injected into the hidden layers of a neural network. This deep integration strategy enables more meaningful interactions between physics-informed features and the evolving system state, extending the guidance beyond raw input augmentation. For example, Pawar et al. (2021a) implement this strategy by proposing a modular physics-guided machine learning framework in which a simplified physical theory is used to generate auxiliary signals. These signals, instead of being used as direct inputs, are inserted into the hidden layers of an LSTM network (paths 2a-4a-3b-4b-5a, **Fig. 2**). In a follow-up work, Pawar et al. (2021b) apply this framework to a model fusion scenario involving reduced-order modeling of fluid flows. They use a Galerkin-based projection method to extract low-dimensional physics-based features, which are then concatenated with the intermediate hidden states of an LSTM network (paths 2a-4a-3b-4b-5a, **Fig. 2**).

### 3.3. Limitations and opportunities of UPGML

Although the literature on UPGML remains scarce in hydrology and other scientific domains, an analysis of its underlying mechanisms and architectural design reveals several limitations.

The first limitation stems from the strong dependence of UPGML on the accuracy of physical model outputs. These outputs are used as features or embedded representations at different stages of the learning process. Consequently, any bias or structural error in the physical model directly contaminates the predictive performance of the ML component (Winstral et al., 2019; Xiang et al., 2021). This reliance undermines the robustness of the entire system, especially in domains where physical models are uncertain or poorly constrained. Future work should focus on integrating mechanisms for model discrepancy correction or adaptive trust-weighting of physical inputs. This would allow the ML component to learn when to rely on the physics and when to override it based on observational evidence.

A second limitation is the computational cost imposed by the physics-based simulator (Ielmini and Milo, 2017; Wang et al., 2022a), which dictates the total runtime of UPGML models. This becomes a major obstacle in applications requiring high-resolution, long-term simulations, such as climate change projections or coupled Earth system modeling. Reducing this burden requires the development of efficient surrogate models (Sun et al., 2023; Xu and Zhang, 2024), reduced-order representations (Chen et al., 2021a; Xiao et al., 2019), or hybrid training strategies where physics-based inputs are precomputed, cached, or selectively updated based on model sensitivity.

Finally, UPGML inherits classical weaknesses from the machine learning paradigm itself. Overfitting, underfitting, and poor generalization remain pressing issues (Aliferis and Simon, 2024; Zhang et al., 2019), particularly when training data are sparse or when the physical signals dominate the learning dynamics. These problems are exacerbated by the additional complexity introduced by combining two modeling paradigms. To address this, new regularization schemes are needed that penalize not just data misfit but also physical inconsistency. Moreover, curriculum learning strategies (Soviany et al., 2022; Wang et al., 2022b) that gradually increase the role of physical guidance during training could improve convergence and generalization.

## 4. Hybrid physics-machine learning

Unlike UPIMLs, where physics is embedded in the training loss, or UPGML where inductive biases are hardwired into the model, hybrid physics–machine learning models preserve a clear separation of roles between physical modules and data-driven components (**Fig. 3**). This is particularly suited to hydrology, where governing processes are often partially known but affected by uncertainty, measurement errors, or unresolved subgrid phenomena. We distinguish three major categories of hybridization: additive learning, physics-embedded machine learning, and submodule replacement (**Fig. 3**). Each category responds to a specific modeling need.

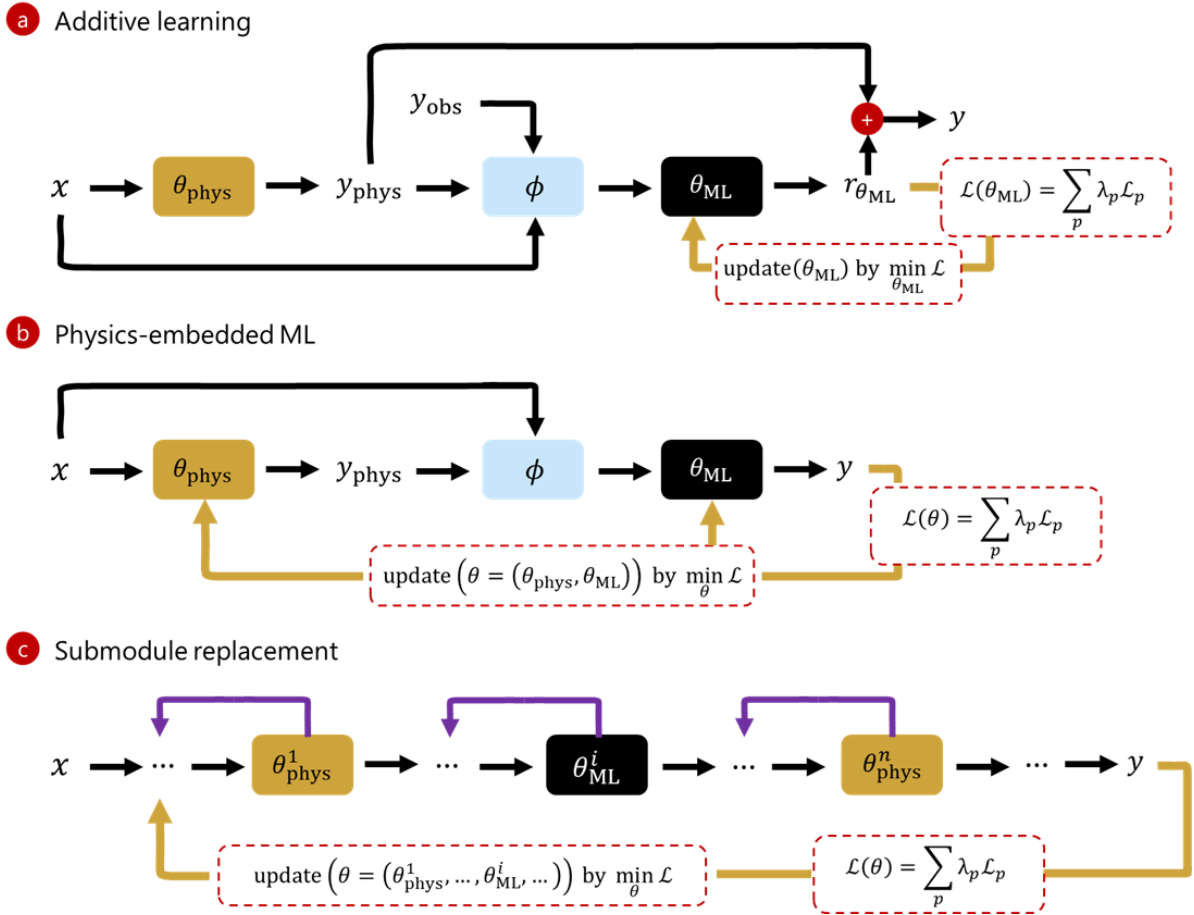

**Fig. 3** General architectures of the three main categories of hybrid physics-machine learning models. Each configuration combines a physics-based module (gold) with machine learning components (black) under different integration strategies. **(a)** Additive learning augments the output of a physics-based model by learning the residual between the observed target and the physics output using a machine learning model. The final prediction is obtained by summing the physics output and the ML-predicted residual. Only the machine learning parameters are updated, using a loss function composed of task-specific terms. **(b)** Physics-embedded machine learning combines physics and machine learning in a differentiable, end-to-end manner. The physics-based model provides intermediate outputs used within or alongside the machine learning model, with both sets of parameters jointly optimized. This structure enables gradient propagation across physics and ML components. **(c)** Submodule replacement substitutes internal components of a physics-based model with machine learning counterparts, while preserving the global model structure. The replaced modules may target poorly understood processes, computational bottlenecks, or empirical relationships. There may also be a cyclical exchange of information between the physics and machine learning modules (purple arrow). $\phi$ (blue module) is a transformation function (e.g., addition, subtraction, concatenation). Learning is achieved through joint optimization over all physical and machine learning submodules.

### 4.1. Additive learning

In additive learning, the machine learning model is trained to predict the residual error between the output of a physics-based model and observed data, with the goal of correcting systematic biases in the physical model's predictions (**Fig. 3a**). Let $x \in \mathbb{R}^d$ denote the input vector (e.g., meteorological forcings) and $y_{\text{obs}} \in \mathbb{R}^m$ be the observed target variable (e.g., streamflow). The physics-based model produces an initial prediction $y_{\text{phys}} = F_{\text{phys}}(x, \theta_{\text{phys}})$, and a machine learning model $r_{\theta_{\text{ML}}} = F_{\text{ML}}(\phi(x, y_{\text{phys}}, y_{\text{obs}}), \theta_{\text{ML}})$ is trained to estimate the residual between $y_{\text{obs}}$ and $y_{\text{phys}}$. $\theta_{\text{phys}}$ denotes either known or calibrated physical parameters; $\theta_{\text{ML}}$ represent the parameters of the machine learning model; $\phi$ is a transformation function such as concatenation, subtraction or even a much more complex one. The final prediction is given by **Eq. 20**.

$$y = y_{\text{phys}} + r_{\theta_{\text{ML}}} \qquad 20$$

For example, Cho and Kim (2022) used the residual error from the physics-based WRF-Hydro model, along with the input forcing data, to train an LSTM model. This LSTM learns to predict the residuals, which are then added to the WRF-Hydro output to produce the final streamflow estimates. Similarly, Konapala et al. (2020) proposed a hybrid modeling approach where an LSTM network was trained to learn the residual errors of a process-based hydrologic model, using five forcing variables as inputs: precipitation, maximum and minimum daily temperature, solar radiation, and vapor pressure. The final streamflow estimates were obtained by adding the LSTM-predicted residuals to the process-based model output, resulting in improved performance across a wide range of catchments, especially where the process-based model performed poorly.

### 4.2. Physics-embedded machine learning

Physics-embedded machine learning (PEML) structurally integrates physics and machine learning in a two-stage modeling design (**Fig. 3b**). In contrast to additive learning or UPGML approaches, these architectures explicitly embed physical knowledge into the internal structure of the ML model and subsequently apply a neural post-processing module that corrects or enhances the outputs of the embedded physics. The two modules are jointly trained or linked to form a coherent prediction pipeline that benefits from both physical interpretability and ML flexibility.

Let $x \in \mathbb{R}^d$ denote the input features (e.g., precipitation, temperature). A first computational block, $F_{\text{phys}}$, applies a differentiable physics-based model, generally implemented as a custom neural network that embeds physical knowledge, to produce an intermediate output. A transformation function $\phi$ then combines the intermediate output with the original inputs; this function can be a simple concatenation or a more complex operation. The resulting representation is subsequently processed by a second machine learning block, $F_{\text{ML}}$, to generate the final prediction $y \in \mathbb{R}^m$ (**Eq. 21**).

$$y = F_{\text{ML}}\left(\phi\left(x, F_{\text{phys}}(x, \theta_{\text{phys}})\right), \theta_{\text{ML}}\right) \qquad 21$$

This strategy has been implemented in recent studies such as Cai et al. (2022), Jiang et al. (2020) or Adombi et al. (2024), where a conceptual bucket-type model is embedded within a machine learning architecture. Jiang et al. (2020) originally proposed this hybrid design by customizing a recurrent layer to embed the conceptual EXP-HYDRO model, followed by a 1D convolutional neural network that acts as a post-processing block to correct the model predictions for streamflow simulation. Similarly, Adombi et al. (2024) applied the same architecture to groundwater level modeling to study the effects of climate change, embedding an HBV or linear model as the physics-based component.

### 4.3. Submodule replacement

Submodule replacement consists in substituting specific components of a differentiable physics-based model with trainable machine learning modules while preserving the broader physical model structure (Chen et al., 2023a; Huynh et al., 2025; Li et al., 2023) (**Fig. 3c**). Let the overall model be a composition of submodules $\mathcal{M}_{\text{full}} = \mathcal{M}_n \circ \cdots \circ \mathcal{M}_1$, where each $\mathcal{M}_i$ corresponds to a hydrological process (e.g., infiltration). In this framework, some submodules $\mathcal{M}_i$ are replaced by machine learning counterparts denoted $\mathcal{M}_{\text{ML}}^{(i)}$. The resulting hybrid model produces the final prediction as given by **Eq. 22**.

$$y = \mathcal{M}_n \circ \cdots \circ \mathcal{M}_{\text{ML}}^{(i)} \circ \cdots \circ \mathcal{M}_1(x), \qquad 22$$

In Li et al. (2023), the EXP-HYDRO conceptual model is embedded within a custom recurrent neural network to obtain a differentiable version of the model. This setup enables the targeted replacement of one to four internal hydrological processes with neural networks, while preserving the overall model structure. The hybrid model was used for streamflow prediction.

Chen et al. (2023a) build upon an eco-hydrological modeling framework originally composed of three physics-based models: WOFOST for crop growth, HYDRUS-1D for vadose zone flow, and MODFLOW for saturated groundwater flow. From this setup, a hybrid version is developed by replacing the MODFLOW component with a radial basis function (RBF) network trained to predict groundwater head based on HYDRUS-1D outflows. The predicted head is then reinjected into HYDRUS-1D as a lower boundary condition. This feedback loop allows HYDRUS-1D to produce updated estimates of key variables such as evapotranspiration and root water uptake, enabling dynamic interaction between machine learning surrogates and the remaining physical components.

### 4.4. Limitations and opportunities of hybrid physics-machine learning

In additive learning, the dependence of the machine learning component on observed data or physics-based model outputs during inference is a major limitation. Since the machine learning model is trained to predict the residual error between the output of the physics-based model and the observations, it generally needs access to these outputs when making predictions. This dependency prevents the ML module from operating independently, making it unsuitable as a stand-alone predictor and limiting the applicability of the overall system mainly to data assimilation or correction tasks for which observations are permanently available. To overcome this limitation, a promising approach is to perform an initial offline training phase during which the ML model learns residuals using both physics-based model outputs and observed data. After this calibration, the parameters of the ML model are "frozen" and the model is then incorporated into a new framework where its inputs depend solely on variables available for operational use, such as the inputs of the physics-based model. This process can be seen as a form of transfer learning or model decoupling (Khoshkalam et al., 2023; Yao et al., 2023) and allows the hybrid model to operate autonomously, enhancing its use in scenarios where continuous observational data is not accessible.

Physics-embedded machine learning requires the physical model to be differentiable, which constrains its applicability in hydrology. Many conventional hydrological models use discrete numerical schemes or legacy codes that are not designed for automatic differentiation (Kim et al., 2008; Lampert and Wu, 2015; Luo et al., 2017). Consequently, it is often impossible to directly compute gradients through these physical models, preventing their seamless integration into an end-to-end learning frameworks. To address this limitation, future research could focus on developing differentiable surrogate models that approximate the original physics-based model or adopt stepwise training strategies that alternate between optimizing physical and machine learning components.

In submodule replacement, identifying which parts of the physical model to replace is non-trivial. It often requires deep knowledge of the physical system and, in practice, is frequently guided by manual trial and error (Li et al., 2023). This process can be time-consuming and prone to subjective bias. Furthermore, inserted ML components may not naturally align with the structural, temporal, or numerical constraints of the surrounding physical architecture, leading to mismatches or degraded global performance. Another critical limitation is that the ML submodules may learn to compensate for deficiencies in the physics-based model in a way that improves predictive accuracy without resolving the underlying physical inconsistency. As a result, structural flaws in physics may be masked rather than corrected. Addressing these challenges calls for principled integration strategies, such as interface-aware design of ML components and the use of physical constraints to regularize learning. In addition, automated diagnostic tools such as sensitivity analysis (Tunkiel et al., 2020; Zhang, 2019) or physics-guided error attribution can support more systematic identification of replacement targets and help ensure that the hybrid model improves both accuracy and physical soundness.

5.  **Physics discovery**

Physics discovery constitutes a distinct subfield within the broader landscape of scientific machine learning in hydrology and focuses on identifying the unknown physical laws themselves, directly from the data. Formally, let $u(s, t) \in \mathbb{R}^d$ denote the state variables of interest at spatial location $s \in \mathcal{S} \subseteq \mathbb{R}^n$ and time $t \in \mathbb{R}_+$. Physics discovery aims to find a dynamic relationship as given by **Eq. 23**.

$$\mathcal{D}[u(s,t)] = F_{\text{disc}}(u(s,t), x(s,t), \theta) \qquad 23$$

where $\mathcal{D}$ is an operator acting on $u$; $x$ denotes exogenous inputs or forcings (e.g., precipitation, vapor pressure); and $F_{\text{disc}}$ is a learnable functional form parameterized by $\theta$ (**Fig. 4**). The nature of the operator $\mathcal{D}$ determines the class of physics discovery formulation considered. In what follows, we classify physics discovery methods into three categories: (1) symbolic regression, (2) stochastic universal partial differential equations and (3) data-driven discovery of conceptual bucket-type model.

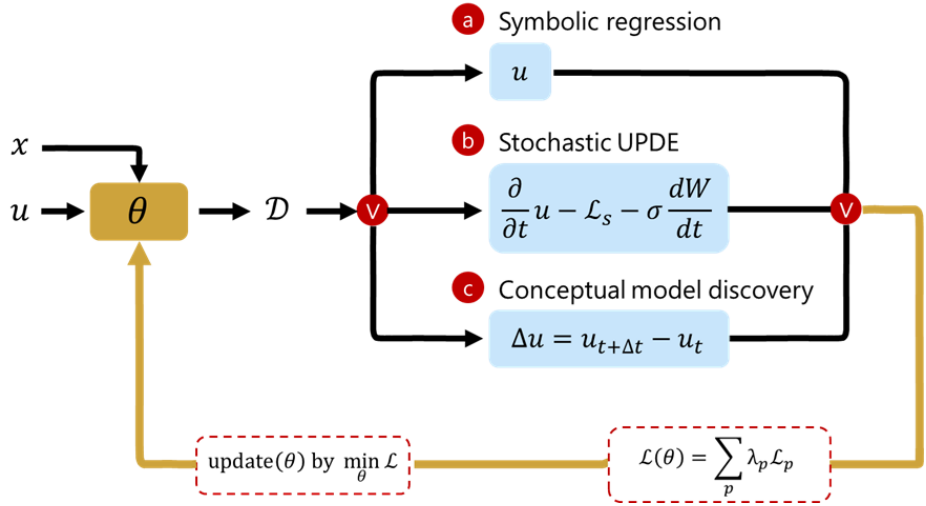

**Fig. 4** General framework of physics discovery, illustrating three alternative modeling strategies: **(a)** symbolic regression aims to uncover explicit algebraic relationships between variables from data, offering interpretable models grounded in closed-form expressions. **(b)** stochastic universal partial differential equations (SUPDEs) aim to model complex physical systems that evolve over space and time by combining mechanistic structure, stochastic variability, and learnable components within a unified differential framework. **(c)** In conceptual model discovery, the system

is formulated in discrete time, often inspired by traditional hydrological models, and seeks to discover model structures or gain insights into dominant hydrological processes. Each approach defines a candidate operator $D$, with selection mediated by a choice operator V.

### 5.1. Symbolic regression

Symbolic regression searches for mathematical equations that best fit the data by combining basic functions (e.g., addition, multiplication, sine, exponential), aiming to discover interpretable and compact expressions that describe the relationships between variables. Within our unified operator-based framework, it corresponds to the most elementary case, where the operator $\mathcal{D}[u(s,t)] = u(s,t) = u$ reduces to an identity operator (**Fig. 4a**). The task consists in discovering a function $F_{\text{disc}}$ such that $u = F_{\text{disc}}(x, \theta)$, where both the structure and the parameters of $F_{\text{disc}}$ are inferred directly from data.

A representative example of this class is the AI Feynman algorithm (Udrescu and Tegmark, 2020), which demonstrates a robust multi-stage pipeline for symbolic regression. It combines neural networks with physics-based heuristics such as dimensional consistency, variable separability, and symmetry constraints to simplify and break down complex relationships into interpretable components. The algorithm iteratively reduces the original regression task $u = F_{\text{disc}}(x, \theta)$ into a simpler one with dimensionless variables $u' = F_{\text{disc}}(x', \theta')$, which is then approximated by low-order polynomials or simpler expressions through brute-force symbolic enumeration. Neural networks are leveraged not for prediction, but to probe properties of the unknown function (e.g., symmetry or additivity), which in turn guides the symbolic search. This approach epitomizes the use of symbolic regression in discovering physically meaningful laws from data with no explicit reliance on temporal or spatial operators.

In hydrology, symbolic regression has also been applied in a more structured context. Klotz et al. (2017) propose a grammar-based symbolic regression method to discover transfer functions that map static catchment attributes to spatially distributed hydrological model parameters. In distributed bucket-type modeling, each spatial cell is governed by a local parameter vector, whose values need to be inferred. Instead of calibrating them directly, the authors define parameterization as a function of catchment attributes through an unknown transfer function $F_{\text{disc}}$. The symbolic regression is used in a two-stage process: first to infer the structure of $F_{\text{disc}}$, and then to calibrate its parameters using the Nelder–Mead simplex algorithm (Gao and Han, 2012; Olsson and and Nelson, 1975) and an augmented Lagrangian approach (Afonso et al., 2011; Lin et al., 2011). This strategy demonstrates how symbolic regression can be embedded into broader physical modeling workflows to automatically induce parametric structure from high-dimensional attribute spaces.

### 5.2. Stochastic universal partial differential equations (SUPDE)

The most general case within our framework is described by stochastic differential equations, extended to include spatial operators to form stochastic universal partial differential equations (SUPDE). SUPDE aims to model complex physical systems that evolve over space and time by combining mechanistic structure, stochastic variability, and learnable components within a unified differential framework. In this framework, the operator $\mathcal{D}[u(s,t)] = \frac{\partial}{\partial t}u(s,t) - \mathcal{L}_s[u] - \sigma(u,x,t)\frac{dW}{dt}$. This operator evolves under both deterministic $\left(\frac{\partial}{\partial t}u(s,t) - \mathcal{L}_s[u]\right)$ and stochastic $\left(\sigma(u,x,t)\frac{dW}{dt}\right)$ influences, where $\mathcal{L}_s$ denotes a spatial differential operator acting on $u$; $\sigma$ represents the noise amplitude function and determines the strength and shape of the uncertainty or noise affecting the system, while $\frac{dW}{dt}$ denotes a white noise process, formalized as the time derivative of a Wiener (Brownian) process $W(t)$, which introduces randomness (stochasticity) into the dynamics (**Fig. 4b**). SUPDE models capture temporal dynamics, spatial processes, and stochastic variability in a single formalism as given by **Eq. 24**.

$$\frac{\partial}{\partial t}u(s,t) = \mu(u,x,\theta) + \sigma(u,x,t)\frac{dW}{dt} \qquad 24$$

where $\mu(u,x,\theta) = F_{\text{disc}}(u,x,\theta) + \mathcal{L}_s[u]$ is the deterministic component of the SUPDE. As stated by ElGazzar and van Gerven (2024), the stochastic term $\sigma$ results from intrinsic or extrinsic factors. When $\sigma$ is constant, extrinsic uncertainties such as unobserved external interactions are modeled. When $\sigma$ is a function of the system state $u$, it captures intrinsic uncertainties, such as uncertainties in the parameters of the drift term. From this general framework, several specific cases can be derived, depending on the nature and structure of the system we seek to model: whether its evolution is driven by deterministic or stochastic processes, whether the governing dynamics governing are partially known or entirely unknown, whether the state is a function of time only or also of space.

### 5.2.1. Special cases of SUPDE

#### 5.2.1.1. Differential equations with known structure

This class assumes that the general structure of the system's dynamics is known, for instance the form of mass balance equations or conservation laws commonly used in hydrology, but some components remain unspecified. These unknowns may include fixed parameters, stochastic uncertainties, or part of the deterministic dynamics. Approximation functions, such as neural networks, are introduced to learn these missing elements directly from data. This includes modeling uncertainty or capturing residual dynamics that are not accounted for by the original formulation. This strategy is precisely illustrated in Bolibar et al. (2023), where a PDE-based glacier dynamics model is used (known system dynamics), but where the diffusivity $D = \theta$ is expressed in the form of a neural network to be learned. Here, $\sigma(u, x, t) = 0$; $\mu(u, x, \theta) = b + \nabla \cdot (D_{NN} \nabla S)$ and $u = H$ represents the ice thickness. $b$ is the surface mass balance and $S$, the glacier surface.

### 5.2.1.2. Differential equations with unknown structure

This class considers situations where neither the parameters nor the fundamental structure of the system's dynamics are known in advance (ElGazzar and van Gerven, 2024). As a result, both the deterministic and stochastic components are fully represented by flexible function approximators, typically neural networks. This approach corresponds to a fully data-driven, black-box modeling paradigm, enabling the discovery of governing equations directly from observational data without relying on prior mechanistic assumptions (Morrill et al., 2021; Tzen and Raginsky, 2019a). In addition, the explicit form of these governing equations can also be uncovered using symbolic regression methods (Rackauckas et al., 2020). It provides maximum flexibility but requires careful training and regularization to ensure physically meaningful results.

### 5.2.1.3. Neural ordinary differential equations

In neural ordinary differential equations (ODEs), the state variable depends solely on a single independent variable, typically time, and the stochastic component of the SUPDE is absent, as shown in **Eq. 25**.

$$\frac{du}{dt} = \mu(u, x, \theta) \qquad 25$$

In this context, the state variable $u$ may represent the water volume within a hydrological reservoir (e.g., soil moisture), while the function $\mu$ corresponds to the net water fluxes affecting this reservoir, including precipitation inputs, evapotranspiration losses, or lateral exchanges. One or more components of these net fluxes can be modeled using flexible function approximators, such as neural networks. Several applications of neural ODEs are reported in the hydrological literature. Höge et al. (2022) integrate neural network estimators into the EXP-HYDRO conceptual model, replacing some internal flow equations, such as evapotranspiration, with neural networks. Alkaabi et al. (2025) integrate CNN-LSTM modules with neural ODEs to predict runoff from precipitation, using the output state learned by the CNN-LSTM to calculate the next state via an ODE solver, which is then used to calculate the final discharge.

### 5.2.2. Methods to fit SUPDE to data

Fitting SUPDE models to data involves embedding a numerical solver into a differentiable computational graph to enable gradient-based optimization with respect to a predefined loss function. This setup supports two main strategies (ElGazzar and van Gerven, 2024): discretize-then-optimize, which differentiates through all solver steps to compute exact gradients, and optimize-then-discretize, which uses the adjoint method to approximate gradients with fixed memory cost. For non-stochastic differential equations, these approaches are generally sufficient for learning model parameters. In contrast, stochastic formulations often require additional strategies to account for randomness in the dynamics, such as variational inference or adversarial methods (Kidger et al., 2021; Li et al., 2020; Tzen and Raginsky, 2019a; Tzen and Raginsky, 2019b). To maintain numerical accuracy across variable time scales, adaptive solvers are commonly employed during training (Höge et al., 2022; Rackauckas and Nie, 2017).

### 5.3. Discovery of conceptual bucket-type model

Conceptual bucket-type models are simplified representations of hydrological systems in which water is stored in abstract reservoirs such as soil, snow, or groundwater and transferred between them through fluxes (or processes) like infiltration, percolation, runoff, and evapotranspiration (**Fig. 4c**). Traditionally, these models are manually designed by experts who define both the system structure (i.e., the connections between reservoirs) and the governing equations based on empirical knowledge and simplified physical reasoning. Their strength lies in their interpretability, low data requirements, and ability to capture dominant hydrological dynamics. However, their fixed structure, reliance on hand-crafted functional forms, and limited adaptability to local variability constrain their ability to represent complex nonlinear behaviors or generalize across diverse catchments. To address these limitations, data-driven discovery of conceptual models aims to automatically learn model structures and equations from data, while preserving physical interpretability and consistency. Three innovative approaches have been identified in the hydrological literature and are presented below: DeepDiscover, mass-conserving perceptron and deep process learning.

### 5.3.1. DeepDiscover

DeepDiscover is a recent contribution to the data-driven discovery of conceptual bucket-type hydrological models (Adombi, 2025). Unlike traditional approaches that embed fixed, expert-defined structures, DeepDiscover infers model architecture directly from data. This flexibility addresses two key limitations of conventional conceptual modeling: the lack of structural adaptability across diverse catchments, and the reliance on predefined functional forms that constrain the discovery of novel process relationships. The model is implemented as a recurrent neural network composed of a directed sequence of elementary modules, termed DiscoverUnits (**Fig. 5**). Each DiscoverUnit represents a conceptual reservoir and is designed to learn associated hydrological processes in a physically consistent, data-driven manner. It receives input either from meteorological forcings or the output of the preceding DiscoverUnit and produces a set of candidate hydrological processes through a multilayer perceptron (MLP). These outputs are constrained by a causal structure to enforce physically plausible dependencies between inputs and outputs using the causal relationships constraints (CRC) developed in Adombi et al. (2024).

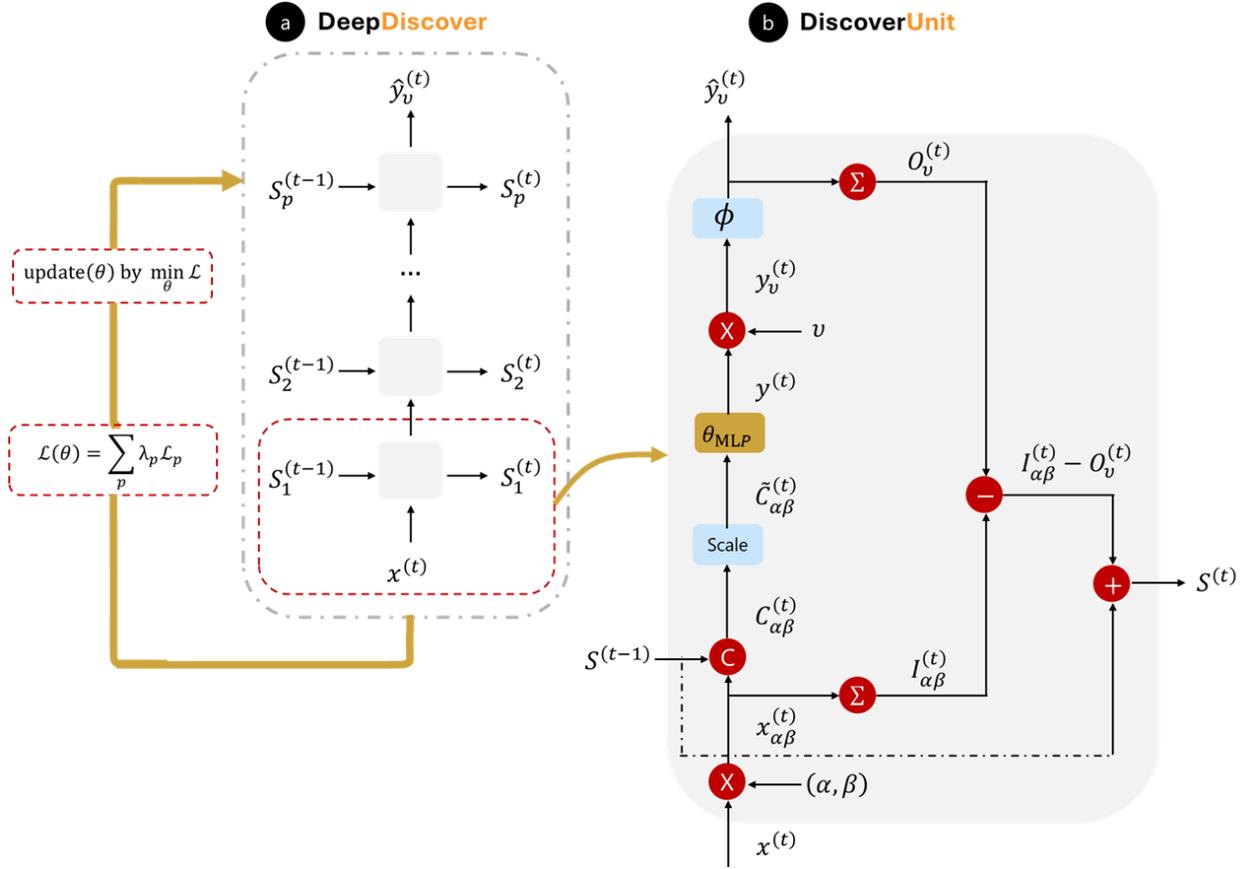

**Fig. 5 (a)** Architecture of DeepDiscover and **(b)** structure of DiscoverUnit. **X** represents element-wise multiplication, **C**, concatenation, and **Σ**, the summation operator. **Scale** is a normalization layer, while MLP is used to generate potential hydrological processes $y^{(t)}$. $\alpha$, $\beta$, and $\upsilon$ are trainable parameter vectors: $\beta$ act as a scaling vector to ensure dimensional homogenization, while $\alpha$ and $\upsilon$ are binary-like vectors with values close to 0 or 1, indicating whether a variable is relevant or not. A filtering mechanism is applied to the estimated potential processes $y^{(t)}$ to obtain effective outputs $y_\upsilon^{(t)}$, which are subsequently corrected into $\hat{y}_\upsilon^{(t)}$ via the function $\phi$ to account for expert knowledge. Modified from Adombi (2025).

To account for known hydrological phenomena such as recession dynamics or thresholds alongside potentially unknown behaviors, a global correction function can optionally be applied to the MLP outputs. While the MLP may learn these effects intrinsically, the correction term provides a mechanism for embedding expert knowledge when necessary. To encourage diversity among the discovered processes, a decorrelation constraint is included in the loss function. Finally, the state $S^{(t)}$ of each DiscoverUnit is updated using a mass balance equation, ensuring consistency with hydrological conservation principles **(Eq. 26)**.

$$\mathcal{D}[u(s,t)] = S^{(t)} - S^{(t-1)} = I_{\alpha\beta}^{(t)} - O_{v}^{(t)} \qquad 26$$

where $I_{\alpha\beta}^{(t)}$ and $O_{v}^{(t)}$ are the sum of the input and output variables at time $t$ respectively. Tested on a streamflow prediction task within a PEML setting, DeepDiscover (DD-PEML) demonstrated superior performance compared to both the traditional EXP-HYDRO conceptual model and its PEML-enhanced variant (EXP-PEML), as well as a data-driven 1D convolutional neural network. Across the CAMELS-US basins, DD-PeML achieved high median scores in terms of Nash–Sutcliffe Efficiency (NSE = 0.68) and Kling–Gupta Efficiency (KGE = 0.70), with low bias and error.

### 5.3.2. Mass-conserving perceptron

The mass-conserving perceptron (MCP) is a neural architecture designed to represent physical systems with strict respect for conservation laws, particularly mass balance (Wang and Gupta, 2024). Implemented as a gated recurrent neural network, MCP provides a compact and interpretable representation of hydrological systems, where each node corresponds to a single dynamic reservoir that maintains an internal storage state and learns how to redistribute incoming water through physically meaningful fluxes (**Fig. 6a**). MCP directly encodes the system's evolution $u_{t+1} = S^{(t+1)}$ through a discrete conservation equation $\mathcal{D}[u(s,t)] = S^{(t+1)} - S^{(t)} = U^{(t)} - O^{(t)} - L^{(t)}$, where $U^{(t)}$ denotes input flux, $O^{(t)}$ the output flux, and $L^{(t)}$ system losses.

These fluxes are learned through parameterized gates constrained in [0, 1]: $U^{(t)} = G_U^{(t)} \cdot U^{(t)}$, $O^{(t)} = G_O^{(t)} \cdot O^{(t)}$, and $L^{(t)} = G_L^{(t)} \cdot L^{(t)}$. The remaining fraction $G_R^{(t)} = 1 - G_O^{(t)} - G_L^{(t)}$ defines a remember gate, controlling how much of the previous state is retained over time. This structure enables MCP to serve two core purposes in scientific machine learning: (i) to provide physically plausible, low-dimensional representations of complex system dynamics, and (ii) to facilitate hypothesis testing through progressive model construction (Clark et al., 2008).

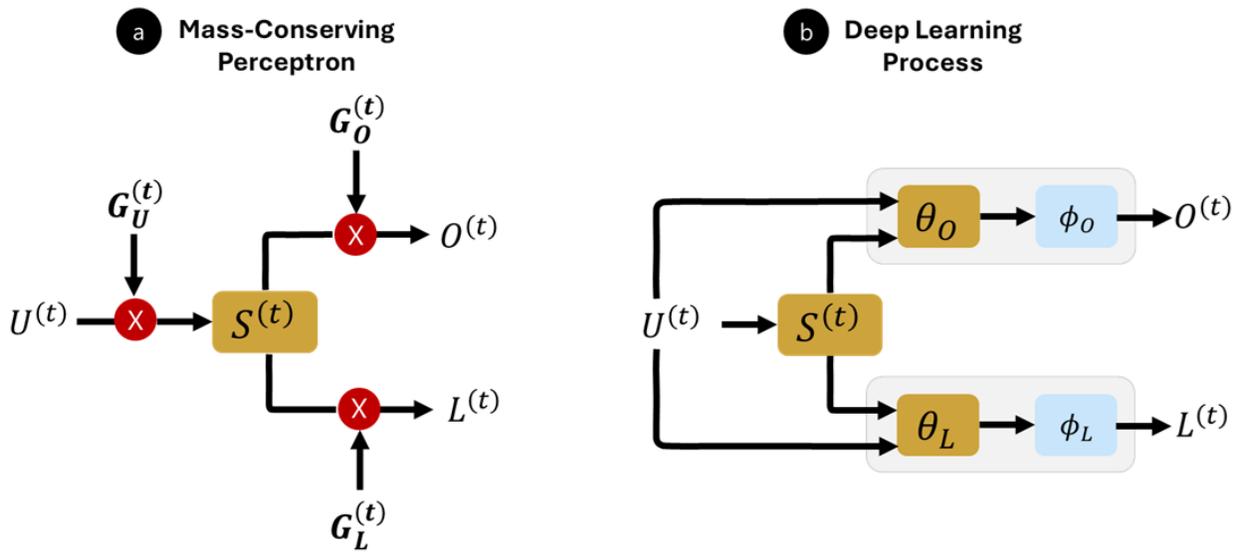

**Fig. 6** Illustration of a simple **(a)** mass-conserving perceptron (MCP) and **(b)** deep learning process (DPL). Both are recurrent neural networks. Each gate $G_i^{(t)}$ is constrained in the interval [0, 1]. $\theta_O$ and $\theta_L$ represent the parameters of a multilayer perceptron whose outputs are transformed using the processing functions $\phi_O$ and $\phi_L$ to produce the output $O^{(t)}$ and loss $L^{(t)}$ fluxes respectively. $S$ is the state of the recurrent neural network.

Unlike DeepDiscover, the model structure and number of output processes are manually defined by the user. In practice, MCP has demonstrated strong performance even in minimal configurations. Applied to the Leaf River Basin, a single-node MCP model achieved high predictive skill (median yearly KGE ≥ 0.85; worst year KGE ≥ 0.50), while capturing the basin's storage dynamics in a physically interpretable way.

### 5.3.3. Deep process learning

Deep process learning (DPL) integrates intuitive physical reasoning directly into neural network architectures to jointly support interpretability and predictive skill (He et al., 2024). **Fig. 6b** is a generic representation of a basic DPL. Unlike DeepDiscover, the model structure and number of output processes are manually defined by the user. For example, the DPL-H4 model implements a two-reservoir system representing snow and soil reserves with three runoff components. It uses a recurrent state-space formulation where the hidden state $u_t = [S_{\text{snow}}, S_{\text{soil}}]$ evolves according to $u_{t+1} = F(x_t, u_t, \theta)$ and $y_t = G(x_t, u_t, \theta)$ with $x_t$ representing meteorological forcings (e.g. precipitation and temperature), $y_t$, the simulated runoff, $F$ defines the state transition function, describing how internal hydrological states evolve over time, $G$ serves as the output mapping, translating the latent states and inputs into an observable output, such as streamflow at the catchment outlet.. To illustrate how intuitive physical reasoning is integrated into the recurrent neural network, consider snow processes. These processes are separated from total precipitation by means of a learnable temperature threshold $\tau = F_T(x_t, \theta) \times \varsigma$ where $F_T$ is a multilayer perceptron parameterized on $\theta$, $\varsigma$ is a user-defined scaling factor, and snow is given by $P_{\text{snow}} = P - P_{\text{rain}}$ where $P_{\text{rain}} = P \odot \frac{\tanh[(T-\tau) \times \xi] + 1}{2}$ with $T$ represents the average temperature, $\xi$ is a value large enough to force the tanh function to reach its limit values (-1,1) and $\odot$ is the element-wise multiplication. DPL directly encodes the system's evolution $u_{t+1} = S^{(t+1)}$ through a discrete conservation equation $\mathcal{D}[u(s,t)] = S^{(t+1)} - S^{(t)} = U^{(t)} - O^{(t)} - L^{(t)}$, where $U^{(t)}$ denotes input flux, $O^{(t)}$ the output flux, and $L^{(t)}$ system losses eventually.

Trained end-to-end using observed runoff data, DPL-H4 achieved high predictive accuracy across CAMELS-US basins while offering interpretable, physically grounded internal dynamics (He et al., 2024).

### 5.4. Limitations and opportunities of physics discovery

In symbolic regression, the sensitivity of the method to noise and data sparsity is a fundamental limitation, often leading to overfitting or the identification of spurious relationships that lack physical validity. Moreover, the combinatorial explosion of the search space with increasing expression complexity renders the discovery of non-trivial equations computationally demanding and potentially unstable (Kahlmeyer et al., 2024; Kammerer et al., 2020). Paradoxically, symbolic regression may even fail to recover relatively simple governing equations, as highlighted by Makke and Chawla (2024), exposing the fragility of current discovery algorithms under realistic data constraints. Despite its theoretical appeal, symbolic regression remains largely absent from the hydrological literature, with most applications confined to synthetic benchmarks.

SUPDEs also pose epistemological challenges, as model identifiability is a key issue. Due to the expressive nature of the stochastic terms, multiple formulations may capture the observed data equally well. This is not inherently problematic from a predictive standpoint, as such flexibility helps account for uncertainties and improves prediction accuracy (Costa et al., 2023; Psaros et al., 2023). However, when the goal shifts to discovering interpretable dynamics, this non-uniqueness becomes a concern. It may obscure the underlying mechanisms and lead to competing hypotheses that are difficult to distinguish or validate. Moreover, SUPDEs could benefit from dense, high-resolution spatio-temporal datasets and involve learning pipelines that can be computationally demanding and sensitive to choices of initialization and hyperparameters. In hydrology, where data may be sparse, noisy or irregularly distributed, these factors can complicate practical deployment (Nyeko, 2015; Yanto et al., 2017).

Conceptual model discovery approaches also face notable methodological challenges. In models such as DeepDiscover, which infer hydrological structure directly from data, a central limitation arises from the mechanism used to disentangle competing candidate processes. When the decorrelation strategy is weak or insufficiently calibrated, the model may converge toward redundant or overly similar process representations. This redundancy undermines the objective of structural discovery, limiting the model's ability to reveal novel hydrological behaviors. Future research should prioritize the development of robust decorrelation techniques to enhance the distinctiveness and scientific value of the inferred processes.

Architectures such as MCP and DPL rely on manually defined model structures, including the number of reservoirs and the nature of the flows between them. This design choice therefore incorporates the modeler's prior assumptions into the architecture. This can introduce a structural bias, limiting the model's ability to discover new interactions or generalize to various catchments. Furthermore, since it is possible to add as many candidate hydrological processes as desired, during training such models may, as in the case of DeepDiscover, tend to generate redundant or overly similar process representations. Having been developed recently, these conceptual model discovery approaches still lack sufficient evaluation across diverse hydrological settings. Thorough testing is crucial to uncover hidden challenges and unlock their full potential for advancing science and practical applications.

## 6. Conclusion

Scientific machine learning is gaining traction as a promising framework for advancing hydrological modeling, but within each methodological family, diverse approaches have emerged in isolation, often without being framed as instances of a broader, unified framework. This review proposed a unified structure for each of these families: unified physics-informed machine learning (UPIML), unified physics-guided machine learning (UPGML), hybrid physics-machine learning models, and data-driven physics discovery. For UPIML approaches, key limitations include high computational cost and limited generalizability under changing conditions. In the case of UPGML, the strong dependence on the accuracy of physics-based model outputs and the high cost of simulation could reduce robustness and scalability. Hybrid physics-machine learning approaches offer integration flexibility but may often suffer from inference dependency, non-differentiable physical components, and subjective choices in model coupling. Physics discovery methods hold promises for uncovering new equations and conceptual model structures but may face significant challenges related to noise sensitivity, model identifiability, and sometimes interpretability, especially in data-scarce settings. Since these approaches have only recently been introduced in hydrology, systematic evaluation across diverse real-world contexts remains essential to reveal hidden limitations and fully realize their scientific and operational potential.

**Author contributions**

**AVDP Adombi:** conceptualization, writing-original draft; writing-review & editing.

**Data availability**

Data sharing is not applicable to this article as no new data were created or analyzed in this study.